\documentclass[12pt]{article}

\usepackage{amsmath,amsfonts,epsfig,color,latexsym}

\newcommand{\be}{\begin{equation}}
\newcommand{\ee}{\end{equation}}
\newcommand{\bea}{\begin{eqnarray}}
\newcommand{\eea}{\end{eqnarray}}

\let\newsection=\section
\renewcommand{\section}{\setcounter{equation}{0}\newsection}

\begin{document}

\begin{flushright}
hep-th/0410124\\
BROWN-HET-1424
\end{flushright}
\vskip.5in

\begin{center}

{\LARGE\bf  On high energy scattering inside gravitational backgrounds
 }
\vskip 1in
\centerline{\Large Horatiu Nastase}
\vskip .5in

\end{center}
\centerline{\large Brown University}
\centerline{\large Providence, RI, 02912, USA}

\vskip 1in

\begin{abstract}

{\large We analyze the high energy scattering inside gravitational 
backgrounds using 't Hooft's formalism. The scattering is equivalent to 
geodesic shifts accross Aichelburg-Sexl waves inside the gravitational 
backgrounds. We find solutions for A-S waves inside various backgrounds 
and analyze them.
}

\end{abstract}

\newpage

\section{Introduction}

PP waves have proven to be an important tool in studying string theory
\cite{bmn,eg,kn}. One 
reason for this is that certain pp waves are exact solutions of string 
theory, as proven in \cite{hos}. In particular, the gravitational 
shockwave of Aichelburg and Sexl \cite{as}
was obtained by boosting to the speed of light
the Schwarzschild black hole while keeping its mass fixed, thus finding 
the gravitational field of an idealized massless particle. It was used as 
a toy model for black hole calculations \cite{dp}. 

But the A-S solution also describes high energy scattering of particles. 
At high enough energies, 't Hooft proved that one can consider a massless
particle is just described by the shockwave. At intermediate energies, a 
scattered particle is described just by geodesics in the A-S wave, but at 
higher energies each massless particle creates its own shockwave. This fact 
was used in \cite{eg} and \cite{kn} to analyze black hole creation in  
high energy scattering in models with low fundamental scale. But another 
application of this idea is to study gauge theory scattering from scattering 
in a gravity dual, a la Polchinski-Strassler \cite{ps}, as was done in 
\cite{kntwo}. In that case, a simple cut-off AdS model was used, but we would 
need more complicated backgrounds to describe more accurately gauge theories.

For these reasons, it is important to find solutions for shockwaves in more
general gravitational backgrounds, and analyze their 't Hooft scattering 
properties. 't Hooft scattering in AdS was also analyzed (using schockwave
solutions) in \cite{aho}. 
We will look at warped compactifications and solutions related 
to black hole backgrounds. We will also use a algorithmic methods for putting 
A-S shockwaves inside other backgrounds, obtained from flat space via a 
T duality, twisting and coordinate change, a procedure know as ``Melvinizing''
(see \cite{ht,akiw}).

The paper is organized as follows. In section 2 we describe 't Hooft scattering
in general gravitational backgrounds. In section 3 we find A-S type solutions 
in warped compactification backgrounds, we reanalyze the backgrounds given 
in \cite{dh}, and find shockwaves inside the maximally supersymmetric pp wave
solution, the noncommutative field theory and the null brane. In section 4 
we look at 't Hooft scattering inside AdS-type backgrounds, the null brane 
and the pp wave and in section 5 we conclude.

\section{'t Hooft scattering and black hole creation in general gravitational
backgrounds}

The Aichelburg-Sexl shockwave \cite{as}
\be
ds^2 =-dx^+dx^-+(dx^+)^2\delta(x^+) \Phi(x^i) +\sum_{i=1}^{d-2} (dx_i)^2
\ee
has as a source a massless particle of momentum p (``photon''), with 
\be
T_{++}= p \delta ^{d-2}(x^i)\delta (x^+)
\ee
In flat space, the Einstein equation $R_{++}= 8\pi G T_{++}$ implies
\be
 \partial _i^2 \Phi (x^i)=-16\pi G p\delta^{d-2}(x^i)
\ee
($\Phi$ is harmonic with source).

't Hooft \cite{thooft,dh} has argued that one can describe the scattering 
of two massless particles at energies close to (but under) the 
Planck scale, ($m_{1,2}\ll M_P, Gs\sim 1$, yet $Gs<1$) as follows.
Particle two creates a 
massless shockwave of momentum $p_{\mu}^{(2)}$ and particle one follows a 
massless geodesic in that metric. 

It is found in \cite{dh}
that geodesics going along u at fixed v are straight except at 
u=0 where there is a discontinuity
\be
\Delta v =\Phi =-4Gp \;ln \frac{\rho^2}{l_{Pl}^2}
\ee

The second effect is a ``refraction'' 
(or gravitational deflection, rather), where the angles $\alpha$ and $\beta$
made by the incoming 
and outgoing waves with the plane $u=0$ at an impact parameter $\rho=b$ from 
the origin in transverse space satisfy 
\be
cot \alpha +cot\beta =\frac{4Gp}{b}
\ee
(here p is the momentum of the photon creating the A-S wave), and at small 
deflection angles (near normal to the plane of the wave) we have 
\be
\Delta \theta\simeq \frac{4Gp}{b}
\ee

The S matrix of the process 
corresponds to a gravitational Rutherford scattering (single graviton 
being exchanged) as follows.

A plane wave $e^{ip\cdot x}$ incident on the A-S metric will be suject 
to the above shift $\Delta v$ and then the scattering amplitude is the 
Fourier transform of this wavefunction, basically 
\be
S=e^{ip_v \Delta v}\equiv e^{ip_-\Phi}
\ee
followed by the impact parameter transform
\be
i{\cal A}= \int\frac{d^{D-2}\vec{b}}{(2\pi)^{D-2}}e^{i\vec{q}\vec{b}}
(e^{i\delta (b,s)}-1)
\ee
with $\delta (b,s)=p_{-}^{(1)}\Phi (b)$. In the original 't Hooft calculation
in \cite{thooft}, in flat D=4, one gets
\be
{\cal A}=
\frac{1}{4\pi} \delta (k_-^{(1)}
-p_-^{(1)})\frac{\Gamma (1-iGs)}{\Gamma(iGs)}
[\frac{4}{(\tilde{p}-\tilde{k})^2}]^{1-iGs}= \delta(k_+^{(1)}-p_+^{(1)})U(s,t)
\ee
and then the differential cross-section is 
\be
\frac{d\sigma}{d^2k}=\frac{4}{s}
\frac{d\sigma}{d\Omega}= 4\pi^2 |U(s,t)|^2= \frac{4}{t^2}(Gs)^2
\ee
In complete generality though, we have \cite{kn}
\be
i{\cal A}= \frac{\Omega_{D-4}\Gamma(\frac{D-3}{2})\sqrt{\pi}2^{\frac{D-4}{2}}}
{(2\pi)^{D-2} q^{D-2}}\int _0^{\infty}dz z^{\frac{D-2}{2}}(e^{i\delta (z=qb,s)}
-1)J_{\frac{D-4}{2}}(z)
\ee

At energies higher than the Planck scale, black holes will start forming. 
The cross section for black hole formation was conjectured in \cite{gt,dl}
(see also earlier work in \cite{bf}) to be the geometric
area of the horizon of a black hole with mass $\sqrt{s}$. In \cite{eg}, 
a framework was developped for analyzing the black hole formation in 
more detail. Based on an observation by 't Hooft \cite{thooft}, the 
scattering was described by the collision of two A-S schockwaves. 
One can prove the formation of trapped surfaces in the A-S collision, even 
though one doesn't know the metric in the interaction region, just the 
single A-S metric. In \cite{kn}, the formalism was extended to arbitrary 
gravitational backgrounds and dimensionality. 

We will not describe it here, but we just note that black hole formation 
in a general background can also be analyzed by using the A-S shockwave inside 
background solutions, so it is important to see under what conditions can 
we get such solutions.

\section{Aichelburg-Sexl waves inside gravitational \\
backgrounds}

{\em Explicit A-S solutions of the Einstein equations}

The first example of an A-S shockwave in a gravitational background was found
by \cite{dh}. They studied the backgrounds of the type 
\be
ds^2=2A(u,v)du dv + g(u, v) h_{ij}(x^i) dx^i dx^j
\label{type}
\ee
The laplacean in this background is 
\be
\Delta_{(d+2)}=\frac{1}{\sqrt{|g|}}\partial_{\mu}g^{\mu\nu} \sqrt{|g|}\partial_
{\nu}= \frac{2}{A}\partial_u\partial_v +\frac{1}{g}\Delta_h+\frac{d}{2Ag}
((\partial_u g)\partial_v +(\partial_v g)\partial_u)
\ee
Then \cite{dh} went on to put a A-S shockwave inside this background by 
finding out the condition to have a geodesic 
shift $\Delta v=f$ at u=0 (which 
as we saw is one way of having an A-S shockwave). The condition was 
\be
\frac{A}{g} \Delta_h f-\frac{2f}{g}g_{,uv}= 32 \pi p A^2 \delta (\rho)
\;{\rm if}\;\;\partial_v A|_{u=0} =0,\;\;\partial_v|_{u=0} g=0
\ee

We will try to reproduce and generalize this shortly, but let's move to a 
different kind of background.  First in \cite{emparan} and later in 
\cite{kn} and \cite{kntwo}, explicit solutions were found for A-S shockwaves 
inside of warped backgrounds of AdS type (1-brane RS model, AdS and 2-brane 
RS model). Let us therefore treat this case in some generality and see 
how the previous solutions appear as particular cases. We will take the 
most general warped background with a shockwave on it,
\be
ds^2 = f(\vec{y}) (-du \;dv +h_{uu} du^2+h_{ij}(\vec{x}) dx^i dx^j)+G_{\mu\nu}
(\vec{y}) 
dy^{\mu} dy^{\nu}
\ee
where $h_{uu}=\Phi(\vec{x},\vec{y})\delta(u)$. 

After a somewhat long calculation, one can prove that the perturbation 
$h_{uu}$ appears only in 
\be
{\Gamma^v}_{uc}=-\partial_c h_{uu}, 
\;\;c=i,\mu;\;\;{\Gamma^i}_{uu}=-\frac{1}{2} h^{ij}\partial_j
h_{uu};\;\; \Gamma^{\mu}_{uu}=-\frac{1}{2} g^{\mu\nu} \partial_{\nu}(f h_{uu})
\ee
and calculating the Ricci tensor
\be
R_{ab}=\partial_c\Gamma^c_{ab}-\frac{1}{2}\partial_a\partial_b ln (g) +
\frac{1}{2}\Gamma^c_{ab}\partial_c ln (g)-\Gamma^c_{ad}\Gamma^d_{cb}
\ee
we find that only $R_{uu}$ contains $h_{uu}$, and it appears linearly,
thus when the background 
($h_{uu}=0$) Einstein equations are satisfied, we have 
\be
R_{uu}= -\frac{1}{2} \frac{1}{\sqrt{h}}\partial_i (h^{ij} \sqrt{h} \partial_j 
h_{uu})-\frac{1}{2}\frac{f}{\sqrt{f^{2+n}G}}\partial_{\mu} 
(G^{\mu\nu} \sqrt{f^{2+n}G}\partial_{\nu}h_{uu})+\frac{1}{2}g_{uu}[
(\partial_{\mu} ln (f))^2-\frac{\Delta_G(f)}{f}]
\ee
But the last term is proportional to $g_{uu}$ and thus can be taken to be 
contributing to the background equation of motion (we have to remember that 
in the Einstein equations, the energy momentum tensor changes when we change 
the metric, for instance if it contains a term proportional to $g_{\mu\nu}$
as is the case here). Thus we have 
\be
R_{uu}= -\frac{f}{2}\Delta h_{uu}+{\rm linear \;in\; }g_{uu}+{\rm background}
\ee
We also note that $g^{uu}=0$, thus there is no $h_{uu}$ contribution in R.
Thus when the background equation is satisfied, the Einstein equation is 
\be
-\frac{f}{2} \Delta_{(x,y)} h_{uu}(x,y)= 8\pi G t_{uu}
\ee
where $t_{uu}$ is the energy-momentum tensor of the shockwave source, that is,
a massless particle:
\be
t_{uu}= p \delta (u) \delta ^n(x^i) \delta^m(y^{\mu})
\ee
so that the Einstein equation reduces to the Poisson equation for $\Phi$ 
in the background:
\be
f(\vec{y}) \Delta_{(x,y)}\Phi (x,y)= -16\pi G p \delta ^{n}(x^i)
\delta ^m(y^{\mu})
\ee

We can check that this encompases the cases analyzed in 
\cite{emparan,kn,kntwo}. For completeness, we will reproduce the solutions 
here. For the 1-brane RS model \cite{emparan}, 
\be
ds^2=dy^2 +e^{-2|y|/l}(-dudv+dx^idx^i+h_{uu}(u, x^i, y)du^2)
\ee
where 
\be
\Phi(\vec{x},y)= \frac{4G_{d+1}}{(2\pi)^{(d-4)/2}}p \frac{e^{d|y|/(2l)}}
{r^{(d-4)
/2}}\int_0^{\infty}dq q^{(d-4)/2}J_{(d-4)/2}(qr)\frac{K_{d/2}(e^{|y|/l}lq)}{
K_{d/2-1}(lq)}
\ee
For the AdS case,
\be
ds^2=dy^2 +e^{-2y/l}(-dudv+dx^idx^i+h_{uu}(u, x^i, y)du^2)
\ee
where 
\bea
\Phi(\vec{x},y)&=& \frac{8G_{d+1}l}{(2\pi)^{\frac{d-4}{2}}}p 
\frac{e^{\frac{dy}{2l}}e^{\frac{4-d}{2l}y_0}
}{r^{\frac{d-4}{2}}}\int_0^{\infty}dq 
q^{\frac{d-2}{2}}J_{\frac{d-4}{2}}(qr) K_{d/2}(e^{y/l}lq)I_{d/2}(e^{y_0/l}lq)
\;\;\;y>y_0\nonumber\\
&=&\frac{8G_{d+1}l}{(2\pi)^{\frac{d-4}{2}}}p 
\frac{e^{\frac{dy}{2l}}e^{\frac{4-d}{2l}y_0}
}{r^{\frac{d-4}{2}}}\int_0^{\infty}dq 
q^{\frac{d-2}{2}}J_{\frac{d-4}{2}}(qr) I_{d/2}(e^{y/l}lq)K_{d/2}(e^{y_0/l}lq)
\;\;\;y<y_0\nonumber\\&&
\eea
and finally on the IR brane of the RS model, we have 
\be
ds^2=dy^2 +e^{2|y|/l}(-dudv+dx^idx^i+h_{uu}(u, x^i, y)du^2)
\ee
where
\be
\Phi(\vec{x},y)= \frac{4G_{d+1}}{(2\pi)^{(d-4)/2}}p \frac{e^{-d|y|/(2l)}}
{r^{(d-4)/2}}\int_0^{\infty}dq 
q^{(d-4)/2}J_{\frac{d-4}{2}}(qr)\frac{I_{d/2}(e^{-|y|/l}lq)}{
I_{d/2-1}(lq)}
\ee
But we see that we can put A-S metrics in general warped backgrounds just by
solving the Poisson equation in the corresponding background. 

We should also note that \cite{lm} has also found an explicit A-S solution 
inside $AdS_3\times S_3$, rotating on the sphere, and smeared over the 
wavefront (so that it doesn't have a $\delta (u)$, but rather is 
u-independent).

Coming back to the Dray- 't Hooft construction, let us take the metric
\be
ds^2 = -A(u,v) du \; dv +g(u,v) [ du^2 h_{uu} +h_{ij} dx^i dx ^j]
\label{dthooft}
\ee
where $h_{uu}=\delta (u) \Phi(x^i)$. Then the nonzero Christoffel symbols are
\bea
&& \Gamma^u_{uu}=\frac{\partial_u A}{A} +\frac{(\partial_v g)h_{uu}}{A};
\;\;\Gamma^u_{ij}=\frac{\partial_v g}{A}h_{ij} \nonumber\\
&& \Gamma^v_{uu}= -\frac{1}{A} (h_{uu}  \partial_u g -\frac{gh_{uu}}{A}
\partial_u A -2 \frac{g(h_{uu})^2}{A} \partial_v g ); \;\;
\Gamma^v_{uv}= -\frac{1}{A} h_{uu} \partial_v g; \nonumber\\&&
\Gamma^v_{ui}=-\frac{g}{A} \partial_i h_{uu}; 
\;\;\Gamma^v_{vv}=\frac{\partial_v
A}{A}; \;\;\Gamma^v_{ij} =\frac{\partial_u g}{A}h_{ij} +\frac{2g}{A^2}h_{uu}
h_{ij} \partial_v g; \nonumber\\
&&\Gamma^i_{uu}=-\frac{\partial^ih_{uu}}{2}; \;\;\Gamma^i_{uk}=\frac{\partial_u
g}{2g}\delta^i_k; \;\;\Gamma^i_{vk}=\frac{\partial_v g}{2g}\delta^i_k;
\;\;\Gamma^i_{jk}=\bar{\Gamma}^i_{jk}
\eea

It is unfortunately quite difficult to calculate the Ricci tensor in general.
But if we restrict to the case of $\partial_v g|_{u=0}=0, \partial_v A|_{
u=0}=0$, as 't Hooft did, we can calculate it, and get, when the background 
Einstein equation is satisfied (keep only $h_{uu}$ terms)
\bea
R_{uu}&=& -\frac{1}{2} \frac{1}{\sqrt{h}} (\partial_i h^{ij}\sqrt{h} \partial_j
h_{uu})+\frac{2g}{A^2}h_{uu}^2 \partial_v^2 g +\frac{g}{A^2}h_{uu} \partial_v 
\partial_u A\nonumber\\&=& -\frac{g\Delta h_{uu}}{2}
+\frac{2g}{A^2}h_{uu}^2 \partial_v^2 g +\frac{g}{A^2}h_{uu} \partial_v 
\partial_u A\nonumber\\
R_{uv}&=&-\frac{1}{A}h_{uu} \partial_v^2 g \nonumber\\\
R_{ij} &=& +\frac{2g}{A^2}h_{uu} h_{ij} \partial_v^2 g 
\label{curva}
\eea
so we see that we actually also need $\partial_v^2 g|_{u=0}=0, 
\partial_v \partial_u A|_{u=0}=0$. 

But $h_{uu}= \delta (u) \Phi\equiv \delta (u) Af/g$ so that we match with 
the calculation of \cite{dh}. And then 
\bea
&& g\Delta_g (\frac{Af}{g}) = A g \Delta _g (\frac{f}{g}) +\frac{2g}{ Ag} 
\partial_u \partial_v A +\frac{f}{gA}(-2+\frac{d}{2})((\partial_u g)\partial_v
A+ (\partial_v g)\partial _u A)\nonumber\\&&
Ag \Delta _g (\frac{f}{g}) =-\frac{2f \partial_u \partial_v g}{g}+\frac
{A\Delta f}{g}+(4-d)\frac{ f(\partial_u g)\partial_v g}{g^2}
\eea
so that when $\partial_v g|_{u=0}=0, \partial_v A|_{u=0}=0$, as well as 
$\partial_v^2 g|_{u=0}=0, \partial_v \partial_u A|_{u=0}=0$, we have 
\be
R_{uu}= -\frac{1}{2}\delta (u) [ \frac{A}{g} \Delta_h f-\frac{2f}{g}g_{,uv}]
\ee
and we indeed get the condition in \cite{dh}.

We note therefore that we can put A-S shockwaves inside general metrics 
of the type in (\ref{dthooft}), but the solution will not come from a 
simple solution of the Poisson equation. We should also note that the 
condition imposed by \cite{dh} is more restrictive, they imposed that one 
should be able to find a coodinate transformation such that we just glue 
two backgrounds along u=0 with a shift $\Delta v=f$, as it happens in 
flat space. This is not guaranteed in a general background. 

We can also understand why we got the simple result (\ref{curva}). If  
$\partial_v g|_{u=0}=0, \partial_v A|_{u=0}=0$, the metric can be written 
as 
\be
ds^2 = g[-(\frac{A}{g} du ) dv + du^2 \delta (u) \Phi +h_{ij} dx^i dx^j]
\ee
and the $(\frac{A}{g} du )$ is just a trivial change of scale, thus the 
metric is of a similar type to the warped background metric. 

Finally, the interest of \cite{dh} in the backgrounds (\ref{dthooft})
is that one can put the Schwarzschild black hole in this form, using 
Kruskal-Szekeres coordinates, 
\bea
&&ds^2=-32 \frac{m^3}{r}e^{-r/2m}dudv +r^2(d\theta^2+\sin ^2 \theta d\phi^2)
\nonumber\\
&&uv \equiv -(r/2m-1)e^{r/2m}
\eea
and get 
\be
f(\theta, \phi)= k \int_0^{\infty}\frac{\sqrt{1/2}cos (\sqrt{3}s/2)}{(\cosh s
-\cos \theta)^{1/2}}ds
\ee

{\em Boosting black hole solutions in backgrounds}

Another way of putting A-S shockwaves inside gravitational backgrounds 
would be to mirror the way Aichelburg and Sexl found their solution. 
Namely, to boost a solution corresponding to a black hole inside a 
gravitational background to the speed of light, while keeping the mass 
parameter fixed. This method, although algorithmically well defined, is 
in practice quite difficult. We will only show a simple example in the 
following, as a check for the A-S inside pp wave solution.

{\em ``Melvinizing''}

One can put an A-S shockwave inside a background also by a procedure 
dubbed ``Melvinizing'' that encompasses several T duality and twist 
examples \cite{akiw} and \cite{ht}. 
It consists of a series of steps performed on 
a supergravity solution possesing an isometry. This procedure was 
used in \cite{aki} for finding a black string solution inside the maximally 
supersymmetric pp wave. 

{\em Review}

The black string inside pp wave solution was obtained using the following 
steps.

1)Write the black string solution as 
\be
ds^2=(1-f(r))dt^2+dy^2-dt^2+\frac{1}{f(r)}dr^2 +r^2d\Omega_7^2
\ee

2)Boost along y: $y=cosh\gamma y'+sinh \gamma t'; t=sinh \gamma y'+cosh \gamma
t'$. 

3)T dualize along y (or $y'$, rather). 

4)Twist the rotation of $S^7$ along y. This amounts to replacing 
\be
d\Omega_7^2\rightarrow d\Omega_7^2+\alpha \sigma dy + \alpha ^2 dy^2
\ee

where

\be
\frac{r^2\sigma}{2}=x_1dx_2-x_2dx_1+x_3dx_4-x_4dx_3+x_5dx_6-x_6dx_5+x_7dx_8
-x_8dx_7
\ee

5)T dualize back along y. 

6)Boost back (with $-\gamma$) along y. 

7)Take the limit $\gamma\rightarrow \infty, \beta=\alpha e^{\gamma}/2=$
fixed. 

One then gets 
\bea
&&ds_{str}^2=-\frac{f(r)(1+\beta^2r^2)}{k(r)}dt^2-\frac{2\beta^2r^2f(r)}{k(r)}
dt dy +(1-\frac{\beta^2 r^2}{k(r)})dy^2\nonumber\\
&&+\frac{dr^2}{f(r)}+r^2d\Omega_7^2-
\frac{\beta^2r^4(1-f(r))}{4k(r)}\sigma^2\nonumber\\
&&e^{\phi}=\frac{1}{\sqrt{k(r)}};\;\;\;k=1+\frac{\beta^2M}{r^4}\nonumber\\
&&B=\frac{\beta r^2}{2k(r)}(f(r)dt+dy)\wedge \sigma
\eea

The solution obtained has the right limits, i.e. for $\beta\rightarrow 0$ one 
gets back the black string solution, and for $M\rightarrow 0$ one is left 
with just the pp wave
\be
ds^2=2du dv -\beta^2 r^2 du^2+r^2d\Omega_7^2
\ee

If one aplies the above procedure ('Melvinizing') to Minkowski space, 
one gets just the pp wave, as one can check. In this case, steps 2 and 3 
are trivial (don't modify the solution), so one can start directly with step
4.

{\em Application: Putting A-S shockwaves inside the pp wave}

Apply this now to an A-S string,
\be
ds^2=2dudv+dy^2-H(r,u)du^2+dr^2+r^2d\Omega^2;\;\;\;H(r,u)=h(r)\delta(u) 
\ee
with boosting along y, as before. If we perform steps 2 and 3, we would get a 
delta function in the denominator. But as we saw, for flat space is wasn't 
necessary, so we will start again at step 4, and see that we get sensible 
results.
One can easily check that one gets 
\be
ds^2=2dudv -du^2[\beta^2r^2 +\tilde{p}\tilde{h}(r)\delta(u)]+dx^2+dr^2
+r^2d\Omega_6^2
\label{assuperp}
\ee
where $h(r)=p\tilde{h}(r); pe^{\gamma}/2=\tilde{p}$=fixed.

For the twisting rotation, one can use only an even number of coordinates, so 
now (in cartesian coordinates for the sphere) $x_1+ix_2\rightarrow e^{i\alpha
y} (x_1+i x_2), x_3+ix_4 \rightarrow e^{i\alpha y} (x_3+ix_4), x_5+ix_6
\rightarrow e^{i\alpha y} (x_5+ix_6), x_7\rightarrow x_7$. Again we have  
 $d\Omega_6^2\rightarrow d\Omega_6^2+ 
\alpha \sigma dy +\alpha^2dy^2$, but now $\sigma$ is defined only for 
coordinates $x_1$ through $x_6$. 

This solution is a slightly trivial superposition of the pp wave and the 
A-S wave. 

Another type of superposition can be obtained by the following procedure:
1)Take the A-S metric as before:
\be
ds^2=-dt^2+dx^2-h(r)\delta(x+t)(d(x+t))^2 +dr^2  + r^2d\Omega_6^2 +dy^2
\ee
2)Boost on x by a $\tilde{\gamma}$ and then boost on y by $\gamma$, with 
$\tilde{\gamma}\gg \gamma\gg 1$. The point of boosting first with 
$\tilde{\gamma}$ is so that the $\gamma$ boost doesn't affect the t in 
the $\delta(x+t)$ term. The $\gamma$ boost doesn't do anything now, but
the $-\gamma $ boost will at the end (same as for pure Minkowski space). 

3)The T duality on y is trivial.

4)The twist acts as before. 5)The T duality on y and 6)The boost on y 
by $-\gamma$ works as in flat space.

7)After the limit, we get 
\be
ds^2=-dt^2+dx^2+dy^2-\beta^2 r^2(dy\mp dt)^2+dr^2+r^2d\Omega_6^2-h(r)
\delta(x\pm t)(d(x\pm t))^2
\ee
That is, a superposition where the pp wave and the A-S wave are boosted on 
different spatial directions. This solution should be what one gets if 
one boosts the pp black string in an x direction. The actual calculation 
is too complicated and will not be done. This is also presumably what one will 
get if one takes the pp wave limit of the AdS-A-S wave (the A-S is in AdS, 
but the pp wave limit boost on the sphere of $AdS\times S$).

Instead, a boost of the pp black string in y gives (\ref{assuperp}). We will 
analyze only the $(t,y)$ part of the metric, the rest will become trivial 
($dr^2+r^2d\Omega_7^2$) in the limit. Then we boost
\be
ds_2^2=(1-\frac{f(r)}{k(r)})dt^2 -dt^2+dy^2 -\frac{\beta^2 r^2}{k(r)}
(f(r)dt^2+2f(r)dt dy +dy^2)
\ee
The trick is that now we have to take not only $M e^{2\gamma}=p$=fixed, 
but also $\beta e^{\gamma}=\tilde{\beta}$=fixed. Then the metric becomes 
\be
ds_2^2=2dudv +du^2 (-\tilde{\beta}^2r^2 +\frac{p}{r^6})
\ee
which is just (\ref{assuperp}), smeared in the u direction (so that 
$\tilde{p}\delta (u)\rightarrow p$) and unsmeared in the x direction.
One can also check that this solution is what one obtains by boosting 
the AdS-Schwarzschild solution times a sphere in the sphere direction and 
``unsmearing'' the solution.\footnote{I thank Aki Hashimoto for pointing 
this out to me and showing me how it works}. This is expected, since the 
AdS times sphere solution becomes the pp wave by boosting, and a Schwarzschild
black hole becomes A-S by boosting.

{\em Putting A-S inside noncommutative field theory.}

Another example of the same ``Melvinizing'' procedure is obtaining the 
NC field theory from flat space. Indeed,
NC field theory is obtained from flat space by the follwing twisting
procedure:

1)Apply twist on $dy^2+dz^2$ as: $y\rightarrow y , z\rightarrow z+\theta y$,
then

2)T duality on y, giving 
\be
ds^2= \frac{dy^2+dz^2}{1+\theta^2};\;\;\; B_{zy} =\frac{\theta}{1+\theta^2};
\;\;\; e^{\phi}=\frac{e^{\phi_0}}{\sqrt{1+\theta^2}}
\ee

3)Then one converts to open string variables using 
\be
G^{\mu\nu}+\theta^{\mu\nu}=(\frac{1}{g+B})_{\mu\nu}
\ee
and obtains 
\be
ds^2=dy^2+dz^2;\;\;\; \theta_{zy}=-\theta
\ee
for the variables felt by the open string. It is understood that everywhere 
we should replace $\theta \rightarrow \theta/\alpha ', B\Rightarrow \alpha '
B, \theta^{\mu\nu}\rightarrow \theta^{\mu\nu}/\alpha '$ and the Seiberg-Witten
limit of NCFT is satisfied by taking $\alpha '\rightarrow 0$. This 
sequence doesn't change the final result.

We would like to mirror this procedure in order to put an A-S solution inside 
the NC field theory.

Unfortunately, again we cannot quite follow the steps used to get an A-S 
inside NCFT. If we take the A-S string 
\be
ds^2=-dt^2+dz^2 +dy^2-h(r) \delta (z+t)(dz +dt)^2+d\vec{r}^2
\ee
and twist $z\rightarrow z+\theta y$ (if we would twist the other way around, 
we couldn't make the T duality, since we would not have an isometry, due to 
the delta function's dependence on the T duality coordinate). 
But still we cannot make the T duality in y before redefining
$t=\tilde{t}-\theta y$, since we have to put the isometry in the canonic 
form before T dualizing. Finally, after the T duality on $\tilde{y}$, we get
\bea
&&ds^2=-dt^2(1+\theta^2)+dz^2(1-\theta^2)+dy^2-h(r) \delta (z+t)(dz+dt)^2 
+d\vec{r}^2\nonumber\\
&& B_{zy}=B_{ty}=\theta
\label{ncas}
\eea
Now we can't be so cavalier anymore about the $\alpha '$ behaviour. We still 
need to replace $\theta \rightarrow \theta/\alpha ', B\rightarrow \alpha ' B$,
but then the Seiberg-Witten limit for NCFT ($\alpha '\rightarrow 0, g_{ij}
\sim \alpha '^2, B_{ij}, G^{ij}, \theta^{ij}$ invariant) 
is not satisfied anymore.  We can fix the metric behaviour by rescaling 
the coordinates $(z,t)\rightarrow \alpha'^2 (z,t);\; y\rightarrow \alpha ' y$,
but we can't make $B_{ij}$ invariant. 

Moreover, if we go to open string variables we get a delta function in 
the denominator. 
\bea
&&\partial^2\equiv G^{ij}\partial_i\partial_j
= \frac{1}{1+\theta^2(\theta^2-2h\delta(z+t))}[-\partial_t^2+\partial_z
^2+h\delta(z+t)
(\partial_t-\partial_z)^2\nonumber\\&&
+2\theta^2\partial_t \partial_z+
(1-\theta^4-2\theta^2h\delta(z+t))\partial_y^2]\nonumber\\
&&\theta^{ty}=\frac{\theta(1-\theta^2)}{1+\theta^2(\theta^2-2h\delta(z+t))}
\nonumber\\
&&\theta^{zy}=-\frac{\theta(1+\theta^2)}{1+\theta^2(\theta^2-2h\delta(z+t))}
\eea
If we do the sequence of rescalings mentioned, we can't make even the 
open string metric invariant, and we are still left with the delta function 
in the denominator. There seems therefore to be no good way of putting an 
A-S schockwave inside NCFT, at most we were able to derive (\ref{ncas}).

{\em A-S inside the null-brane and spacetime-varying NCFT}

A last example of the ``Melvinizing'' procedure is the one used in \cite{hs}, 
(see also \cite{clo} for a generalization of that work)
that we will quickly review.

Starting with flat space,
\be
ds^2 = -2dx^+ dx^- +dx^2+dz^2
\ee
and defining new coordinates by the (singular) transformation 
\bea
&&\hat{x}^+=x^+,\;\;\; \hat{x}^-=x^--\frac{zx}{R}+\frac{z^2 x^+}{2R^2}
\nonumber\\
&&\hat{x}=x-\frac{zx^+}{R},\;\;\;\hat{z}=\frac{z}{R}
\eea
one gets the null-brane background 
\be
ds^2=-2d\hat{x}^+d\hat{x}^-+d\hat{x}^2 +((\hat{x}^+)^2+R^2)d\hat{z}^2+ 
2(\hat{x}^+ d\hat{x}-\hat{x}d\hat{x}^+)d\hat{z}
\label{nullbrane}
\ee
The further coordinate transformation 
\be
\hat{x}=y^+y,\;\;\;\hat{x}^+=y^+,\;\;\;\hat{x}^-=y^-+\frac{y^+ y^2}{2}
\ee
brings it to 
\be 
ds^2=-2dy^+dy^-+R^2 d\hat{z}^2+{y^+}^2 (dy+d\hat{z})^2
\label{nullbrtoo}
\ee
and after the further redefinition $\bar{y}=y+\hat{z}, \hat{z}=\bar{z}/R$
one gets 
\be
ds^2=-2dy^+ dy^- +d\bar{z}^2 +{y^+}^2 d\bar{y}^2
\label{bgr}
\ee
which is related to the original (flat) coordinates by
\be
x^+=y^+,\;\;x=y^+\bar{y}, \;\;z=\bar{z},\;\;x^-=y^-+\frac{y^+\bar{y}^2}{2}
\ee
In \cite{hs}, one starts with the last metric, with a D3 brane in it, 
T dualize on z to a D2 in the same background, twist and redefine the 
coordinates by going back to $y, \hat{z}$, thus reaching a D2 in the 
background (\ref{nullbrtoo}), and then T dualize again to a D3. 
The final background, when viewed in open string variables, is a NCFT
with $\theta^{xz}=R x^+,\theta^{-z}=R x$. 

One gets the dual of the NCFT by putting the metric of N D3 branes in 
(\ref{bgr}) (which is just flat space in different coordinates, so it 
is trivial to write down), following the same steps and taking the 
near-horizon (decoupling) limit at the end. Thus one finds a gravity 
dual to a spacetime-dependent NCFT.

Moreover, the NCFT on the D3 is T-dual to D2 branes in the 
``null-brane'' background of (\ref{nullbrtoo}).

So one can try and follow the same procedure and apply it on a A-S wave
\be
ds^2=-2dx^+dx^--h(r)\delta (x^{\pm})(dx^{\pm})^2 +dz^2
\ee
Putting the A-S inside the null-brane background (\ref{nullbrane}) 
thus adds to (\ref{nullbrane}) either
\be
-h(r)\delta (\hat{x}^+)(dx^+)^2
\label{nbsca}
\ee
or
\be
-h(r) \delta(\hat{x}^-+\hat{z}\hat{x}+\frac{\hat{z}^2x^+}{2})
\label{nbscat}
\ee

Putting an A-S inside the background (\ref{nullbrtoo}) is equally easy 
following the steps in \cite{hs} towards the NCFT. One just needs to add 
to (\ref{nullbrtoo}) either
\be
-h(r) \delta(y^+)(dy^+)^2
\label{nbscatt}
\ee
or 
\be
-h(r) \delta (y^-+\frac{y^+(y+\hat{z})^2}{2})(d(y^-+\frac{y^+(y+\hat{z})^2}{2})
)^2
\ee
The last step towards the NCFT, the T duality on $\hat{z}$, is not modified 
in the first case, but is impossible in the last case (because is no longer 
an isometry).

\section{Scattering examples}

In this section we will give examples of 't Hooft scattering 
in gravitational backgrounds using the 
shockwave solutions described in this paper. 

We can apply the formalism described in section 2 only if the shockwave 
produces just a shift $\Delta v$ at u=0. This was the case in flat space, 
for any shockwave profile. But it is not guaranteed that the same will 
hold true in general. However, in the cases studied in this paper that still 
applies. For the Dray-'t Hooft case, the solution was found by explicitly 
requiring such a fact. For the warped background case, on the brane (at 
$\vec{y}=0$), we still have flat space, so we can apply the 4d 't Hooft 
formalism: we just have a different shockwave profile, but we apply the 
4d scattering rules. For the $\vec{y}=0$ slice we have therefore proven 
the formalism, as described in section 2. Finally, for the ``Melvinizing'' 
procedure, the background is found by applied a series of steps to flat 
space, for which again the 't Hooft formalism applies. So we have to 
define 't Hooft scattering before ``Melvinizing'' and then apply the 
steps to see what we get in the final background. 

We will start with the A-S solution on the IR brane of the 2-brane RS model
\cite{kntwo}, with 
\be
\Phi = l R_s e^{-2|y|/l} \int_0^{\infty} dq J_0 (qr) \frac{I_2(e^{-|y|/l}lq)}
{I_1(lq)}
\ee
At y=0,
\bea
\Phi &\simeq & R_s \sqrt{\frac{2\pi l}{r}}C_1 e^{-\bar{q}_1r}, \;\;\;
r\gg l\nonumber\\
&\simeq & \frac{lR_s}{r},\;\;\; r \ll l
\eea

At nonzero y and $le^{-|y|/l}/r \gg 1$, 
\be
\Phi\simeq \frac{lR_S}{r}e^{-\frac{3|y|}{2l}}\int _0^{\infty} dz J_0(z)
exp(-\frac{l}{r} z(1-e^{-|y|/l}))= \frac{lR_s e^{-\frac{3|y|}{2l}}}
{\sqrt{ r^2 + l^2 (1-e^{-|y|/l})^2}}
\ee
that reproduces the expected 5d result for $y\ll l$,
\be
\Phi \simeq \frac{lR_s}{\sqrt{r^2 +y^2}}
\ee

At nonzero y and $r \gg l$ we can easily check along the lines of the 
calculation in \cite{kntwo} that 
\be
\Phi \simeq lR_s e^{-2|y|/l}\sqrt{\frac{2\pi l}{r}}C_1(y) e^{-\bar{q}_1r}
\ee
where 
\be
C_1(y) = j_{1,1}^{-1/2} \frac{J_2(j_{1,1}e^{-|y|/l})}{la_{1,1}}
\ee
which however is only true if y is not too large, so that $e^{-|y|/l}\sim 1$.

We turn to the calculation of the scattering amplitude, in the limits when 
we can do the integral. 
Since $\delta = p_-^{(1)}\Phi$ and $p_-^{(1)}p^{(2)}=s/4$, $R_s l p_-^{(1)}
=G_5s$, if $\delta $ is small, we get, in the case $lq e^{-|y|/l}\gg 1$
($l e^{-|y|/l}/r\gg 1$) 
\bea
{\cal A}&\simeq &\frac{AG_5s}{q} e^{-\frac{3|y|}{2l}}\int_0^{\infty} 
dz z J_0(z)\frac{1}{\sqrt{ z^2 + l^2 q^2 (1-e^{-|y|/l})^2}}\nonumber\\
&=& \frac{G_5 s}{2\pi } \frac{s}{\sqrt{t}}e^{-\frac{3|y|}{2l}}exp[
-\sqrt{t}l (1-e^{-|y|/l})]
\eea
(in d=4, $A=1/(2\pi)$)
and the condition of $\delta$ small is $G_4 s e^{-3|y|/(2l)}\ll 1$. 

In the case $lq \ll 1$ ($r\gg l$), $e^{-|y|/l}\sim 1$, we get 
\be
\delta (b,s)= G_5 s e^{-2|y|/l}\sqrt{2\pi lq}C_1(y) \frac{e^{-\bar{q}_1z/q}}
{\sqrt{z}}
\ee
which is always small, so 
\be
{\cal A}\simeq \frac{\sqrt{l}G_5s e^{-2|y|/l}C_1(y)}{(2\bar{q}_1)^{3/2}}
{}_2F_1 (\frac{3}{4}, \frac{5}{4}, a' -\frac{t}{\bar{q}_1^2})
\ee

We will also revisit the case of 't Hooft 
scattering on the UV brane, or the 1-brane RS model, studied by \cite{emparan},
making some small generalizations in the process, namely looking for 
't Hooft scattering near the UV brane.

At nonzero y, and $r \gg l$, we find by expanding (the alternate form of) 
$\Phi$ in \cite{emparan},
\be
\Phi=-R_s[log \frac{r^2}{l^2}-\frac{2l}{\pi}e^{2|y|/l}\int_0^{\infty}
dm K_0(mr)\frac{Y_1(ml)J_2(ml e^{|y|/l})-J_1(ml)Y_2(mle^{|y|/l})}
{J_1^2(ml)+Y_1^2(ml)}]
\ee
that if $e^{|y|/l} $ is not too large, the leading order in both terms 
is unchanged, namely 
\be
\Phi(r,y)
\simeq -4G_4p [log \frac{r^2}{l^2}-\frac{l^2}{r^2}+...]
\ee
which is consistent to what we found in \cite{kn}, namely that $\partial_y
\Phi|_{y=0}=0$ and $\partial^2_y\Phi |_{y=0}= o(l^2/r^4)$ (so that 
the first y correction is of the type $(y^2/l^2) (l^4/r^4)$, with $y^2/l^2$
of order 1). 

At nonzero y and $r\ll l$, we get 
\be
\Phi \simeq \frac{4G_4 l p e^{2|y|/l}}{\sqrt{r^2 +l^2 (e^{|y|/l}-1)^2}}
\ee
so that the amplitude in the $ql \gg 1$ case is 
\be
{\cal A}\simeq
\frac{G_5 s}{2\pi } \frac{s}{\sqrt{t}}e^{\frac{2|y|}{l}}exp[
-\sqrt{t}l (e^{|y|/l}-1)]
\ee
whereas for $ql \ll 1$ we have the same leading result as 't Hooft. 
Emparan \cite{emparan} also analyzed the corrections to the leading 
't Hooft result. 

Finally, let's turn to the analysis of 't Hooft  scattering inside 
backgrounds obtained by ``Melvinizing''.

Let us analyze scattering in the null-brane background (\ref{nullbrane}). 
We take a plane wave in the flat space coordinates,
\be
\Psi_{(1)}= e^{i p\cdot x}
\ee
and apply the coordinate transformation 
\be
x= \hat{x}+\hat{z} \hat{x}^+;\;\; x^+=\hat{x}^+;\;\; z=R\hat{z};\;\; 
x^-=\hat{x}^-+\hat{z}\hat{x}+\frac{\hat{z}^2\hat{x}^+}{2}
\ee
Then 't Hooft scattering is defined by the shift $\Delta x^-=h$, $\Delta
x=\Delta z=0$ at $x^+=0$, which implies that the scattered wavefunction 
for the case defined by (\ref{nbsca}) is
\be
\Psi_{(2)}=\Psi_{(1)}e^{-ip^+h}\equiv \Psi_{(1)}S(s,b)
\ee
where we have defined the S matrix in impact parameter space as the 
ratio of wavefunctions. 

Similarly, for the case defined by (\ref{nbscat}), we have 
\be
\Psi_{(2)}= \Psi_{(1)} e^{-i(p^+\hat{z}^2/2+p^-- p^x \hat{z})h}
\equiv \Psi_{(1)}S(s,b)
\ee

For the background in (\ref{nullbrtoo}), we need to apply on the 
flat space plane wave the coordinate transformation 
\be
x^+=y^+;\;\;\hat{z}=z/R;\;\; x=y^+(y+\hat{z});\;\ 
x^-=y^-+y^+\frac{(y+\hat{z})^2}{2}
\ee
and 't Hooft scattering is defined by a shift $\Delta y^- =\Delta x^-=h$,
$\Delta y=\Delta \hat{z}=0$ at $y^+=0$. We get the scattered wavefunction
\be
\Psi_{(2)}= \Psi_{(1)}e^{-ip^+h}\equiv  \Psi_{(1)}S(s,b)
\ee 

And lastly, the scattering on the pp wave is defined similarly. The 
incoming wavefunction is 
\be
\Psi_{(1)}= e^{ip^+x^-+ip^-x^+-i\beta^2 r^2 p^+x^+ +ip^y y +i \vec{p}_r \vec{r}
}
\ee
and the shift at $x^+=0$ is $\Delta x^-=h$, giving 
\be
\Psi_{(2)}=\Psi_{(1)}e^{ip^+h}\equiv \Psi_{(1)}S(s,b)
\ee

\section{Conclusions}

In this paper we have found solutions for A-S waves inside gravitational 
backgrounds. For warped backgrounds, one needs just to solve the Poisson 
equation in the given background. For the backgrounds in (\ref{dthooft}) 
in general it is complicated to find a solution, but in the case that the
conditions in \cite{dh} are imposed, it is much easier. Shockwaves inside
the maximally supersymmetric pp wave, noncommutative field theory and the 
null brane are slightly trivial to obtain. 

We have calculated the effect of 't Hoooft scattering inside AdS-type 
backgrounds. It is possible to do this in a general warped background 
since the scattering takes place on a brane, thus obeying the 4d rules
(the calculation is basically the same as for 't Hooft, except the wave 
profile is different). 
In the case of \cite{dh}, the solution was found by imposing that there
is a geodesic shift $\Delta v=\Phi$, thus also being able to calculate
't Hooft scattering.
One can also calculate 't Hooft scattering in 
backgrounds related to flat space by a series of steps. We have given as 
examples the pp wave and null brane backgrounds. In that case, the S matrix 
is defined as the ratio of the scattered wave function to the incoming 
wavefunction.

{\bf Acknowledgements} I would like to acknowledge very useful conversations 
with Aki Hashimoto, both on his work and on shockwaves, as well as 
for a critical reading of the manuscript, and to thank the 
UW Madison group for hospitality, allowing me to have these conversations.   
This research was  supported in part by DOE
grant DE-FE0291ER40688-Task A.

\end{document}